\documentclass[aps,prb,reprint,groupedaddress,floatfix]{revtex4-1}

\usepackage{xspace}
\usepackage{graphicx}
\usepackage[nosumlimits]{amsmath}

\newcommand*{\MyChi}{\raisebox{0.35ex}{\( \chi \)}}%

\begin{document}


\title{Inelastic neutron scattering studies of YFeO$_3$}


\author{S. E. Hahn}
\author{A. A. Podlesnyak}
\author{G. Ehlers}
\author{G. E. Granroth}
\affiliation{Quantum Condensed Matter Division, Oak Ridge National Laboratory, Oak Ridge, TN, 37831, USA}

\author{R. S. Fishman}
\affiliation{Materials Science and Technology Division, Oak Ridge National Laboratory, Oak Ridge, TN, 37831, USA}

\author{A. I. Kolesnikov}
\affiliation{Chemical and Engineering Materials Division, Oak Ridge National Laboratory, Oak Ridge, TN, 37831, USA}

\author{E. Pomjakushina}
\author{K. Conder}
\affiliation{Laboratory for Developments and Methods, Paul Scherrer Institut, CH-5232 Villigen-PSI, Switzerland}


\date{\today}

\begin{abstract}
Spin waves in the the rare earth orthorferrite YFeO$_3$ have been studied by inelastic neutron scattering and analyzed with a full four-sublattice model including contributions from both the weak ferromagnetic and hidden antiferromagnetic orders.
Antiferromagnetic (AFM) exchange interactions of $J_1$= -4.23$\pm$0.08 (nearest-neighbors only) or $J_1$ = -4.77$\pm$0.08 meV and  $J_2$ = -0.21$\pm$0.04 meV lead to 
excellent fits for most branches at both low and high energies. An additional branch associated with the hidden antiferromagnetic order was observed. This work paves the way for studies of other materials in this class containing spin reorientation transitions and magnetic rare earth ions.
\end{abstract}


\maketitle

\section{Introduction}
The rare earth orthorferrites \textit{R}FeO$_3$ are an important family of materials whose 
magnetic properties remain a focus of 
considerable research due to promising applications in innovative 
spintronic devices.\cite{Kimel} Furthermore, they contribute to an emerging class of 
materials, i.e., multiferroics with strong magnetoelectric (ME) coupling.\cite{Eerenstein,Tokunaga}
In multiferroic materials, coupling between magnetic and ferroelectric order gives rise to magnetization 
on application of an electric field or to electric polarization on application of a 
magnetic field. Their complex non-collinear structures and magnetic phase transitions are due 
to the combination of the antiferromagnetic (AFM) exchange interaction with the Dzyaloshinsky-Moriya (DM) 
antisymmetric exchange interaction.\cite{Dzyaloshinsky,Moriya}
 
In general materials in the \textit{R}FeO$_3$ family contain two magnetic subsystems consisting of either iron or rare earth ions. 
With decreasing temperature or an applied magnetic field, most of these materials undergo a spin reorientation transition from $\Gamma_4 \left(G_a,F_c\right)$, where the net moment is along 
the \textit{c}-axis, to $\Gamma_2 \left( G_c,F_a\right)$, where the net moment is along the \textit{a}-axis (for notation, see appendix A).\cite{White1969} This transition occurs over
a finite temperature range where the spins rotate continuously in the lower symmetry phase 
$\Gamma_{24} \left(G_{ac},F_{ca}\right)$. No structural change is observed in ErFeO$_3$ and YbFeO$_3$, suggesting 
that this is purely a magnetic transition.\cite{Bazaliy2005,Tsymbal2005} Rotation of the iron moments leads to a change in the magnitude of the 
magnetization on the rare earth subsystem, which must be included in the calculation of the rotation angle
and absolute magnetization. At lower temperatures an additional magnetic transition occurs when the rare earth moments order.   

The nonmagnetic 
yittrium sublattice in YFeO$_3$ enables us to focus only on the magnetic 
interactions of the iron sublattices. The lack of a spin reorientation transition with temperature
considerably simplifies the modeling of spin dynamics and makes YFeO$_3$
a good stepping stone to studying other materials in this class with more complex dynamics.

\begin{figure}[htp]
 \includegraphics{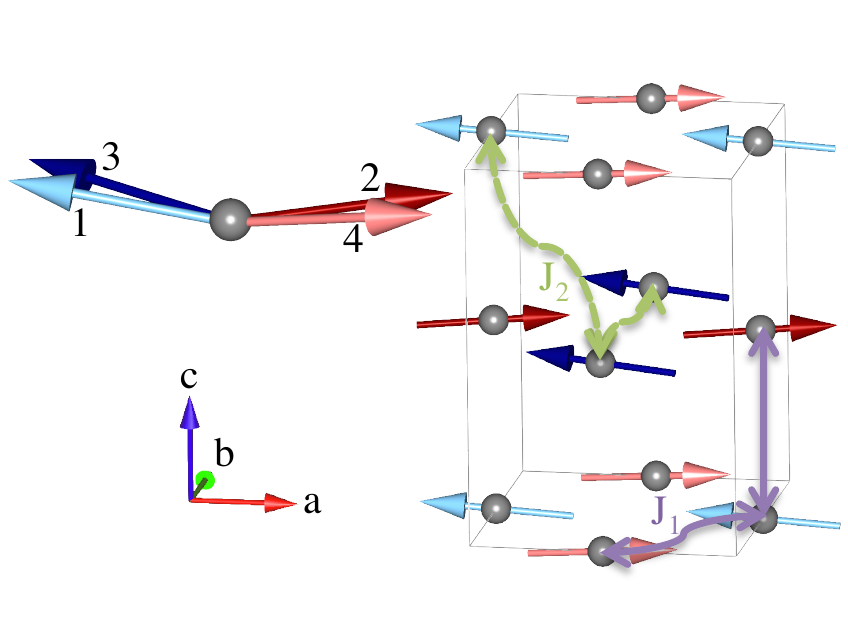}%
 \caption{\label{mag_struct} (color online) Magnetic unit cell of YFeO$_3$, showing only the
 positions of the Fe$^{3+}$ atoms. The four sublattices show weak ferromagnetism 
 and antiferromagnetism along the \textit{c} and \textit{b} directions, respectively. 
 Exchange interactions between nearest ($J_1$) and next-nearest ($J_2$) neighbors
 are shown by the solid purple and dashed green arrows, respectively.}
 \end{figure}

YFeO$_3$ adopts an orthorhombic structure with space group \emph{Pbnm}. 
Below 640K, YFeO$_3$ is a non-collinear antiferromagnet whose four 
Fe$^{3+}$ ions are in the state $\Gamma_4(G_a,F_c,A_b)$, shown in Fig. \ref{mag_struct}. 
The ratio of $A_b/G_a$, which determines 
the canting angle along \textit{b}, was found to be 1.59(7)$\cdot$10$^{-2}$.\cite{Plakhty1983} Values for $F_c/G_a$ range from 8.9$\cdot$10$^{-3}$ to 1.29$\cdot$10$^{-2}$ where the lower values may be due to ferromagnetic impurities. These values set limits on
the canting angle along \textit{c}. 

Spin waves in similar systems TmFeO$_3$ and ErFeO$_3$ were previously 
measured and modeled with a combination of four-sublattice and two-sublattice 
models for the short-wavelength and long-wavelength 
dispersion, respectively.\cite{Shapiro1974} For TmFeO$_3$, the exchange constant 
for nearest neighbors only ($J_2\equiv0$ ) was found to be \hbox{$J_1$ = -4.22 meV}. With next-nearest 
neighbors included, the exchange constants were {$J_1$ = -5.02 meV} and 
\hbox{$J_2$ = -0.324 meV.} A four-sublattice model containing only exchange predicts reasonable energies 
for the observed spin wave branches.   
The two easy-axis anisotropy parameters are approximately equal near the transition and an additional term proportional to the 
fourth power of the spin controls the rotation angle over the temperature range of the transition. 

TbFeO$_3$ has generated renewed interest since measurements in an applied magnetic field found an unusual incommensurate phase with a periodic array
of widely separated domain walls. The ordering of domain walls is due to a long-range force from the exchange 
of magnons propagating through the iron sublattice.\cite{Artyukhin2012} Spin waves in TbFeO$_3$ were previously 
measured and modeled with a four-sublattice 
model containing only exchange interactions.\cite{Gukasov1997} 
In principle, the distortion from cubic symmetry leads to different exchange 
constants for nearest and next-nearest 
neighbors within the \textit{ab} plane and between planes.  Measured 
distances between Fe$^{3+}$ ions are however within 2\% and 6\% of each other for nearest neighbors and next-nearest neighbors, respectively.
In each of these cases, the exchange parameters within the \textit{ab} plane and between planes can be treated as equal. 
With only nearest neighbors, 
$J_1$ = -4.34 meV and with both nearest and next-nearest neighbors the
exchange constants were $J_1$ = -4.95 meV and 
$J_2$ = -0.241 meV. 

In YFeO$_3$, spin waves were measured at 1.4 and 2.2 meV in the long-wavelength limit with Raman scattering at room 
temperature.\cite{White1982} 
A two-sublattice model was then used to obtain estimates for the anisotropy 
constants, defined in Eqn. \ref{ham}, of \mbox{$K_a =  4.6\cdot10^{-3}$ meV}
and \mbox{$K_c= 1.13 \cdot10^{-3}$ meV}. In this work,
we measured spin waves in YFeO$_3$  by inelastic neutron scattering on two different energy scales and analyzed them 
simultaneously with a quantitative model considering contributions from both the weak ferromagnetic and 
hidden antiferromagnetic orders present in the full four-sublattice model.

\section{Experiment}

Polycrystalline YFeO$_3$ was prepared by a solid state reaction. Starting 
materials of Y$_2$O$_3$ and Fe$_2$O$_3$ with 99.99\% purity were 
mixed and ground followed by a heat treatment in air at 1000-1250$^\circ$C 
for at least 70 hours with several intermediate grindings. Phase purity of 
the resulting compound was checked with a conventional x-ray diffractometer. The 
resulting powder was hydrostatically pressed into rods (8 mm 
in diameter and 60 mm in length) and subsequently sintered at 
1400$^\circ$C for 20 hours. 

The crystal growth was carried out using an optical floating zone furnace 
(FZ-T-10000-H-IV-VP-PC, Crystal System Corp., Japan) with four 500W 
halogen lamps as heat sources. The growing conditions were: 
the growth rate was 5 mm/hour, the feeding and seeding rods were rotated 
at about 15 rpm in opposite directions to ensure the liquid's homogeneity and 
an oxygen and argon mixture at 1.5 bar pressure was applied during growth. 
Lattice constants in the \textit{Pbnm} space group were $a=5.282$ \AA, $b=5.596$ \AA, and $c=7.605$ \AA.
The sample was orientated in the (\textit{H0L}) plane for neutron measurements.

Inelastic neutron scattering measurements were done using the 
Cold Neutron Chopper Spectrometer (CNCS)\cite{Ehlers2011} and the 
Fine Resolution Chopper Spectrometer (SEQUOIA)\cite{Granroth2006,Granroth2010} 
at the Spallation Neutron Source (SNS) at Oak Ridge National Laboratory.
The data were collected using fixed incident neutron energies 99.34~meV (SEQUOIA) 
and 3.15~meV (CNCS), which allowed for the measurement of excitations up to 
energy transfers of $\Delta \omega \sim$ 80~meV (SEQUOIA) and 2.5~meV (CNCS).
In these configurations, a full width at half maximum (FWHM) resolution of 5.5~meV (SEQUOIA) and 0.06~meV 
(CNCS) was obtained at the elastic position. The sample was cooled to 4 K on SEQUOIA and base temperature ($<2$ K) on CNCS.
The MantidPlot \cite{MantidPlot} and DAVE \cite{DAVE} software packages
were used for data reduction and analysis.

\section{Theoretical Modeling of Spin Waves}

Our model Hamiltonian, given in Eqn. \ref{ham}, contains isotropic exchange constants $J_1$ and $J_2$ coupling nearest-neighbor 
and next-nearest-neighbor Fe$^{3+}$ spins, two DM antisymmetric exchange constants $D_1$ and $D_2$ 
responsible for the canting along \textit{c} and \textit{b} and two easy-axis anisotropy constants $K_a$ and $K_c$ 
along the \textit{a} and \textit{c} axes. 

\begin{align}
\label{ham}
H=& -J_1  \sum_{\left<i,j\right>}  \boldsymbol{S}_i \cdot \boldsymbol{S}_j  
      -J_2 \sum _{\left<i,j\right>'} \boldsymbol{S}_i \cdot \boldsymbol{S}_j \nonumber  \\
     &  -D_1 \sum_{\boldsymbol{R}_j = \boldsymbol{R}_i + a\left( \hat{x} \pm \hat{y} \right)} \hat{y} \cdot \boldsymbol{S}_i \times \boldsymbol{S}_j  \nonumber \\
     & - D_2 \sum_{\boldsymbol{R}_j = \boldsymbol{R}_i + a\left( \hat{x} \pm \hat{y} \right)}  \hat{z} \cdot \boldsymbol{S}_i \times \boldsymbol{S}_j \nonumber \\
     & -K_a \sum_i \left( \boldsymbol{S}_i^x \right)^2 
      -K_c \sum_i \left( \boldsymbol{S}_i^z \right)^2 
\end{align}

\noindent The DM interaction was only considered among nearest neighbors 
within the \textit{ab} plane, which is the minimum necessary to explain the canting of all four sublattices.
A third DM interaction is possible along $b$ with 
nearest neighbors between planes, but is not needed to describe the spin 
structure and would add additional complexity to our model. 

Each of the four spins are written in spherical coordinates as
\begin{equation}
\boldsymbol{S_i} = S \left(\textrm{sin}\,\theta_i \, \textrm{cos}\, \phi_i, \textrm{sin}\, \theta_i \, \textrm{sin}\, \phi_i, \textrm{cos}\, \theta_i \right)
\end{equation}

\noindent where $S= 5/2$. As a first step in this analysis, one must find the angles associated with the minimum classical energy. By assuming that 
$\theta_i = \theta$ for all sublattices and  $\phi_1 = \pi + \phi$, 
$\phi_2 = \phi$, $\phi_3 = \pi - \phi$, and $\phi_4 = 2\pi - \phi$, the number of independent angles is 
reduced to two.  Assuming small angles, one can linearize the problem and find the expressions in Eqn. \ref{theta_angles} and Eqn. \ref{phi_angles} for $\theta$ and $\phi$ to lowest order as a function of 
$J_1$, $J_2$, $D_1$, $D_2$, $K_a$ , and $K_c$.

\begin{align}
\label{theta_angles}
\theta &= \frac{\pi}{2} +  \frac{2 D_1}{6 J_{1} + K_c - K_a} \\
\label{phi_angles}
\phi &= - \frac{2 D_2}{4 J_{1} - 8 J_{2} -K_a}
\end{align}

\noindent In the above expressions, $\theta \leq \frac{\pi}{2}$ and $\phi > 0$.  Next these expressions were then used to find values for $D_1$ and $D_2$ that produce the experimentally 
determined canting angles. The ratio of $F_c/G_a$ = 1.29$\cdot$10$^{-2}$ and $A_b/G_a$ =  1.59$\cdot$10$^{-2}$ were used to fix the angles $\theta=$ 1.5656  (89.70$^\circ$)  and $\phi=$ 0.0032 (0.18$^\circ$).\cite{Plakhty1983}
 
The inelastic neutron cross section for undamped spin waves is

\begin{align}
S \! \left( \boldsymbol{q}, \omega \right) &= \sum_{\alpha,\beta} 
     \left( \delta_{\alpha \beta} - q_\alpha q_\beta / q^2 \right) S_{\alpha \beta} \! \left( \boldsymbol{q}, \omega \right) \\
     &=  \sum_{n,\alpha} \left[ 1 - (q_\alpha/q)^2 \right] \delta(\omega - \omega_n ( \boldsymbol{q} )) S^{(n)}_{\alpha\alpha} \! \left( \boldsymbol{q} \right)
\end{align}

\noindent where $\alpha$ and $\beta$ are the cartesian directions $x, y, z$ and
$n$ enumerates the individual branches.\cite{Shirane2004} $S_{\alpha \beta} \! \left( \boldsymbol{q}, \omega \right)$ is the spin-spin 
correlation function describing undamped spin waves at low temperature. The spin-spin 
correlation function is diagonal when there is no net moment and antisymmetric 
otherwise, meaning that off-diagonal elements do not contribute to the intensity.
The energies $\omega_n ( \boldsymbol{q} )$ and terms contributing to the scattering intensities 
$S^{(n)}_{\alpha\alpha} \! \left( \boldsymbol{q} \right)$ were solved using the 
1/S formalism outlined in Ref. \onlinecite{Haraldsen2009} and appendix A of Ref. 
\onlinecite{Fishman2013}. For direct comparison with experimental intensities,
the effects of the magnetic form factor, instrumental resolution function, and integration width
were included in our calculations according to appendix B.

 \begin{figure}[htp]
 \includegraphics[width=3.4in]{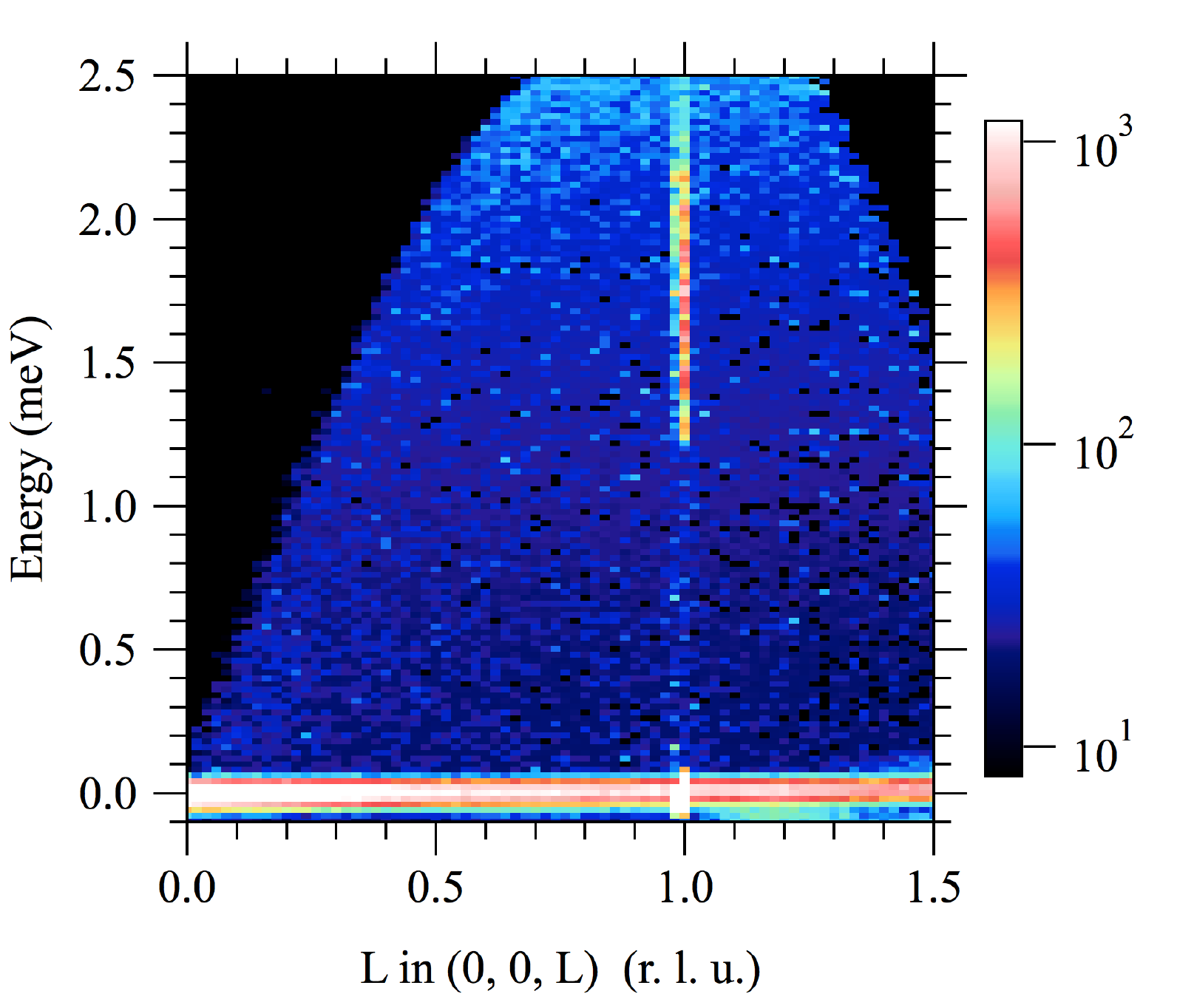}%
 \caption{\label{CNCS_01L}  (color online) Spin wave energy gap in YFeO$_3$ measured by inelastic neutron scattering.}
 \end{figure}
     
\begin{figure}[htpb]
\begin{tabular}{c}
 \includegraphics[]{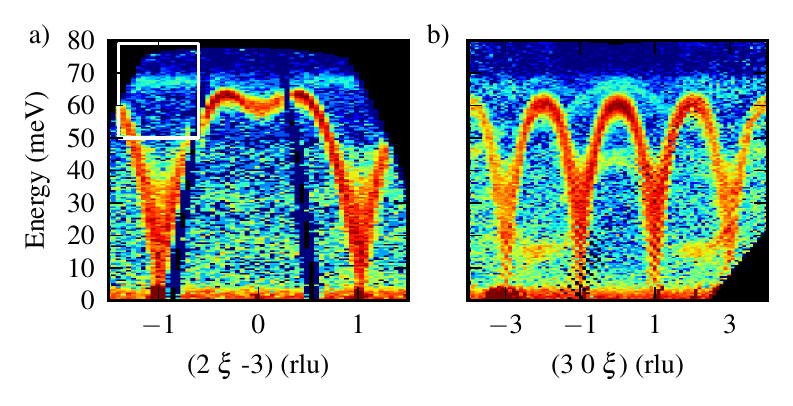}\tabularnewline
 \includegraphics[]{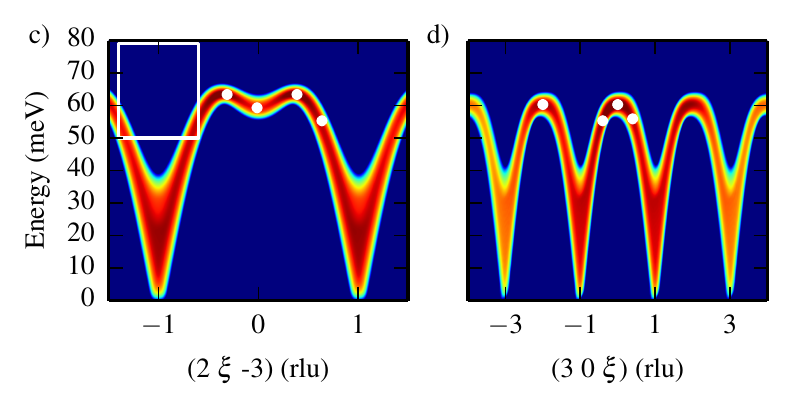}\tabularnewline
  \includegraphics[]{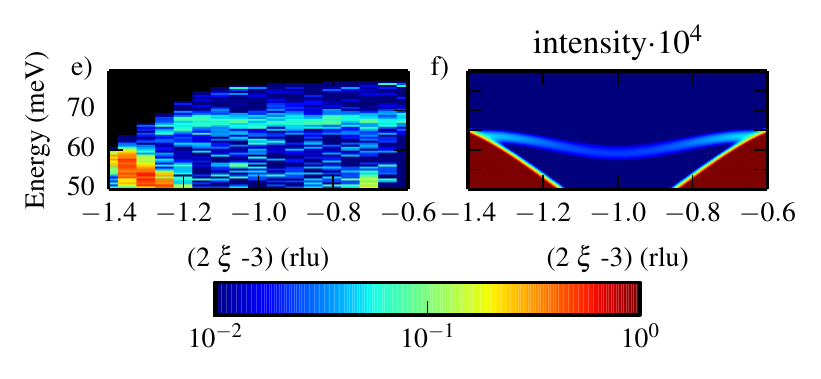}\tabularnewline
\end{tabular}
\caption{\label{YFeO3_30L} (color online) Measured 
spin wave dispersion along a) $(2,\xi,{-3})$ and b) $(3,0,\xi)$ 
and calculated spin wave dispersion along c) $(2,\xi,{-3})$ 
and d) $(3,0,\xi)$. The background contains phonon modes not included in our model. 
e) Enlargement of the region outlined by the white box in a). 
f) Enlargement on the region outlined by the white box in c) and with the intensity 
multiplied by 10$^4$. In all figures, the white dots show energies used for fitting. 
Black pixels show regions where no data was collected.}
\end{figure}

To find the set of parameters that best fits the data, the energy with the highest intensity was taken at eight points in reciprocal 
space that described the shape of the spin wave dispersion. Our model finds two branches with similar energies contributing to 
the highest intensity branch. Energy differences range from 0.8 meV at the zone center to 0.01 meV at the zone boundary. 
These branches are, however, too close in energy to be resolved separately in the cuts shown in 
Figs. \ref{YFeO3_30L}c and \ref{YFeO3_30L}d, so the energy bin with the highest intensity was compared against the average 
of the two energies weighted by their intensities. 

At the zone center we used the observed energies from Ref. \onlinecite{White1982} of 1.4 and 2.2 meV.  The lower value is in good agreement
with measurements from CNCS shown in Fig. \ref{CNCS_01L}, though we were not able to independently verify the frequency of the 
second mode. The variance was estimated by a Gaussian fit to the measured data. Exchange and anisotropy parameters 
$J_1,J_2,K_a$ and $K_c$ were fitting parameters,  $D_1$ and $D_2$ were 
adjusted for each calculation using Eqn. \ref{theta_angles} and the canting angles $\theta$ an $\phi$ remained fixed. 
The NLopt nonlinear-optimization package\cite{NLopt} was used for 
the least squares fitting. Error bars indicate when the reduced $\MyChi^ 2$ increases by 1.0. For $D_1$ and $D_2$ we propagated the
error assuming a 10\% error in the canting angles.

\begin{table*}
\caption{\label{parameters_comparison} Best fit parameters used in this work and 
compared with other work on YFeO$_3$ and similar materials. In the second line only
nearest neighbors were included (J$_2$=0.0). All values are in meV}
 \begin{ruledtabular}
 \begin{tabular}{l c c c c c c c}
Material &  Num. SL &$J_1 $& $J_2$ & $D_1$ & $D_2$ & $K_a$ & $K_c$ \\
YFeO$_3$  & 4 & -4.77$\pm$0.08 &  -0.21$\pm$0.04 & 0.074$\pm0.008$ & 0.028$\pm$0.003 & 0.0055$\pm$0.0002 & 0.00305$\pm$0.0002 \\
& 4 & -4.23$\pm$0.08 & 0.0 & 0.066$\pm$0.007 & 0.028$\pm$0.003 & 0.0063$\pm$0.0002 & 0.0036$\pm$0.0002 \\
YFeO$_3$\cite{White1982} & 2 & -4.96  & 0.0 & 0.11 & & 0.0046 & 0.0011 \\
TmFeO$_3$\cite{Shapiro1974} & 4&  -5.01& -0.32 &  & & &\\
& 4 &-4.22 & 0.0 &  & & &\\
TbFeO$_3$\cite{Gukasov1997} &4& -4.94 & -0.24 &  &  & &\\
&4 & -4.34 & 0.0 &  &  & &\\

 \end{tabular}
 \end{ruledtabular}
\end{table*}

Parameters determined from this fit along with data from similar work on YFeO$_3$ and similar materials 
is given in Tbl. \ref{parameters_comparison}. Values for $J_1$ are considerably lower than those published 
by White et al., possibly because their fit considered only the long wavelength limit. Our results are
similar to those of Shapiro et al.\cite{Shapiro1974} and Gukasov et al.\cite{Gukasov1997} in similar materials. Anisotropy parameters are not equal in 
the 2-sublattice and 4-sublattice models because the hidden canting is absorbed into renormalized anisotropy 
parameters.\cite{Herrmann1964}  Therefore anisotropy parameters should not be directly compared between two and four sublattice models.
  
\section{Discussion}

Overall, excellent fits are obtained for most branches.
Figs. \ref{YFeO3_30L}a and \ref{YFeO3_30L}c show the measured and calculated spin wave dispersion along (2, $\xi$, {-3}).
Both show a dip in frequency and intensity at (2, 0, {-3}) and the integration range and experimental resolution explain the line width.
The intensity around 70 meV near $\xi$ = {-1},1 appears to be magnetic scattering and is not visible in our calculation
on this intensity scale. Fig. \ref{YFeO3_30L}e enlarges the region in Fig. \ref{YFeO3_30L}a near $\xi$ = {-1} outlined by the white box.  Spin wave branches in this region are also reported elsewhere.\cite{Shapiro1974,Gukasov1997} The two-sublattice model does not contain any branches near $\xi$={-1},1 that could explain this intensity. Two-magnon scattering occurs around 120 meV, well above the range of energy transfers we measured.\cite{Koshizuka1988}
A full four-sublattice model doubles the unit cell, leading to zone folding and consequently two additional
branches close to those energies. The hidden antiferromagnetic order gives these additional branches
some intensity, motivating us to model this system with the full four-sublattice model. 

When $\phi=0$ this branch has zero intensity, consistent with zone folding in a supercell. A nonzero value of $\phi$ makes sublattices 
1,3 and 2,4 unequal and gives this branch some intensity. Fig. \ref{YFeO3_30L}f
shows this same region in our calculation, though with the intensity multiplied by $10^5$.  At these small angles, changing $D_2$, and consequently the $\phi$ angle, has the greatest effect
on the intensity of this branch whereas chaining $D_1$, or the $\theta$ angle, has little if any effect. The ratio of the intensities of these two branches is more than four orders of magnitude too weak compared to the measured ratio of 0.07. Small changes in these angles alone are not enough to account for this difference. In addition, measured energies are up to 9 meV higher than what would be expected from zone folding. Quantum fluctuations are missing from this model and may have an effect on these energies and intensities. 

Agreement remains excellent along other directions. Figs. \ref{YFeO3_30L}b and \ref{YFeO3_30L}d show the calculated and measured spin wave dispersion along $\left(3, 0, \xi \right)$.
Calculated energies agree well with the measured values. The integration range and resolution function accounts for the observed 
widths, especially at low energies. Aluminum and phonon scattering have not been subtracted and may 
account for any structure seen in the background.
 
\begin{figure}[htpb]
\begin{tabular}{c}
 \includegraphics[]{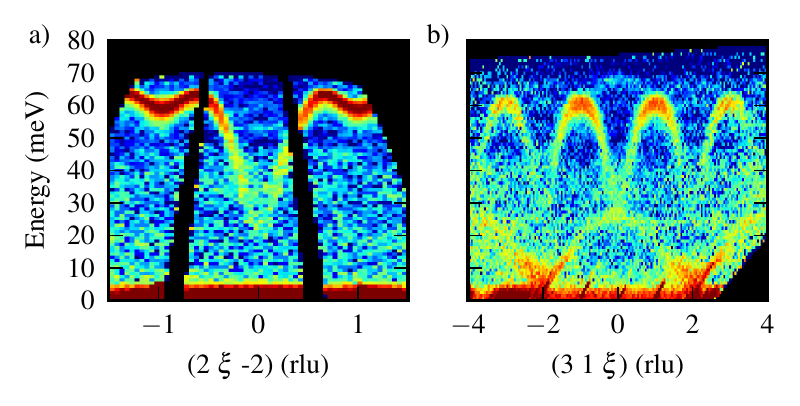}\tabularnewline
 \includegraphics[]{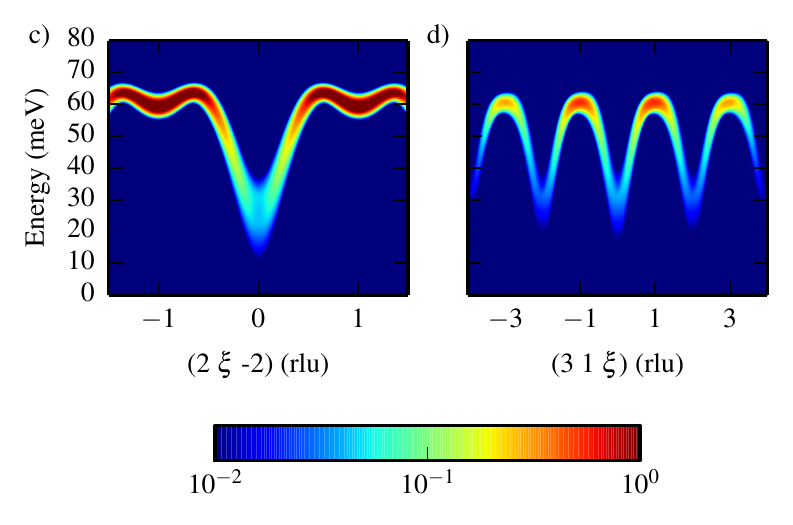}\tabularnewline
\end{tabular}
\caption{ \label{YFeO3_2K-2}  (color online) a) Measured spin wave dispersion along a) (2, $\xi$, {-2}) and b) (3, 1, $\xi$) 
and calculated spin wave dispersion along c) (2, $\xi$, {-2}) and d) (3, 1, $\xi$). The phonon mode with double periodicity was not included in our model. Black pixels show regions where no data was collected.}
\end{figure}

To show how the spin wave intensities depend on the location in reciprocal space, Figs. \ref{YFeO3_2K-2}a and \ref{YFeO3_2K-2}c show the calculated and observed spin wave dispersion along (2, $\xi$, {-2}). Despite identical energies, the intensity is dramatically different from that observed along (2, $\xi$, {-3}) in Fig. \ref{YFeO3_30L}. Along ( 2, $\xi$, {-2}) the intensity approaches the background at low energies and increases dramatically with energy. The change from \textit{L} = {-3} to \textit{L}={-2} changes the reciprocal lattice points from Q-type, $(h,k,l)$ = (even, odd, odd), to O-type, $(h,k,l)$ = (even, even, even).\cite{Shapiro1974}  

Figs. \ref{YFeO3_2K-2}b and \ref{YFeO3_2K-2}d show the calculated and observed spin wave dispersion along $\left(3, 1, \xi \right)$.    In this
direction the intensity also approaches background at low energies and increases dramatically with energy.  The change from \textit{K} = 0 to \textit{K} = 1 also changes the reciprocal lattice points from Q-type, {$(h,k,l)$ = (odd, even, odd)}, to O-type, {$(h,k,l)$ = (odd, odd, even).}

An additional phonon mode is observed below 25 meV with twice the periodicity of the spin wave. The change in periodicity can be explained by the different unit cells corresponding with the crystallographic and magnetic structures. Ignoring small distortions of the yittrium and oxygen atoms from their ideal positions, treating the two iron sublattices as inequivalent atoms doubles the length of the unit cell along \textit{c}. Views of the powder average show non-dispersive modes at 15, 32, and 82 meV. The intensity of the 15 and 32 meV modes increases with higher Q, suggesting phonon excitations. The 82 meV mode was only measured over a very narrow range in 
Q that was insufficient to identify its Q-dependence. 

\section{Conclusion}

In conclusion, the inelastic spin wave spectrum was measured in the rare-earth
orthoferrite YFeO$_3$ and analyzed with a quantitative model considering contributions from both the weak ferromagnetic and hidden antiferromagnetic orders present in the full four-sublattice model. Excellent fits were obtained that agree well with most observed energies and intensities at both high and low energies. In addition, we observe weak magnetic scattering associated with the hidden antiferromagnetic order along \textit{b}. Future work will explore changes in the spin wave spectrum with spin reorientation as well as materials 
where the rare earth also contains magnetic interactions.

\begin{acknowledgments}

We would like to acknowledge helpful conversations with Jason Haraldsen. 
S.E.H. and R.S.F. acknowledge support by the Laboratory's Director's
fund, Oak Ridge National Laboratory. Research at Oak Ridge National LaboratoryÕs Spallation
Neutron Source was supported by the Scientific User Facilities
Division, Office of Basic Energy Sciences, US Department of
Energy.

\end{acknowledgments}

\appendix

\section{Symmetry Analysis}

The magnetic symmetry of the rare earth orthoferrites is described by linear 
combinations of the spins on four sublattices. The linear combination 
$G_{x_i} = - \vec{M}_1 + \vec{M}_2 - \vec{M}_3 + \vec{M}_4$ describes the primary
G-type antiferromagnetic ordering, $F_{x_i} = \vec{M}_1 + \vec{M}_2 + \vec{M}_3 +\vec{M}_4$ 
describes the weak ferromagnetism, and $A_{x_i} = - \vec{M}_1 + \vec{M}_2 + \vec{M}_3 - \vec{M}_4$ 
describes the weak antiferromagnetism. The subscript $x_i$ gives the direction of these vector quantities. 
In YFeO$_3$, the G-type antiferromagnetic ordering is along \textit{a}, the weak ferromagnetism is along \textit{c}, and the 
weak antiferromagnetism is along \textit{b}.

 \section{Resolution Convolution}

For direct comparison with experimental intensities, the effects of the magnetic form factor 
and the instrumental resolution were included in the calculation. The total intensity is given by

\begin{equation}
I\!\left( \boldsymbol{Q}_0, \omega_0 \right)\! = \!\!\int\!\!\!\!\int \!\!F_Q^2\,S\left( \boldsymbol{Q},\omega \right) 
    R\left( \boldsymbol{Q}\! -\!\boldsymbol{Q}_0, \omega \!-\! \omega_0 \right) d\boldsymbol{Q} \,d\omega
\end{equation}

\noindent where $\boldsymbol{Q}=\boldsymbol{q}+\boldsymbol{G}$ differs by the reciprocal lattice vector 
$\boldsymbol{G}$ and may be outside the first Brillouin zone. The Fe$^{3+}$ magnetic form factor results 
in a lower intensity at higher values of $\boldsymbol{Q}$ and can be approximated as 
{$F_Q=j_0\left( Q \right)$}, where $j_0\left(Q\right) = A_0 e^{-a_0s^2} +  B_0 e^{-b_0 s^2}  + C_0 e^{-c_0 s^2} + D_0$
and $s = \mathrm{sin} \theta / \lambda = Q/(4\pi)$. The coefficients are $A_0$ = 0.3972 ($a_0$ = 13.2442),
$B_0$ = 0.6295 ($b_0$ = 4.9034), $C_0$ = -0.0314 ($c_0$ = 0.3496) and $D_0$ = 0.0044 from Ref. \onlinecite{neutrondatabooklet}.

The experiment resolution shape was approximated by a Gaussian encapsulating a simulated resolution volume.  
For various points along the dispersion,  the resolution was calculated using a full model of the incident beam line 
of SEQUOIA\cite{Granroth2006,Granroth2010} followed by a second model that consists of the Resolution Sample 
and Resolution Monitor components. Both simulations were performed using the McStas\cite{lefmann1999mcstas} 
Monte Carlo package.  First, $36\cdot10^{10}$ neutron packets were propagated down the incident beamline 
simulation. Neutron packets that succeeded in making it to the sample position were stored for later use in the 
secondary spectrometer simulation. Next, for each desired value of $\boldsymbol{Q}$ all of the stored neutrons 
from the upstream simulation were sent through the downstream simulation. Results from this second simulation 
provides a probability function of  $t$ and detector pixel that is transformed to $\omega$ and $\boldsymbol{Q}$ 
based on the kinematics of the measurement and the orientation of the crystal\cite{lumsden2005ub}  for several 
points along the dispersion. Projections of these ellipsoids were taken for planes of the data and a two dimensional 
Gaussian was fit around the 50\% level of the observed projection of the distribution.      

In two dimensions the Gaussian function is proportional to $f(x)=\exp\left(-\zeta^T\!\!A\zeta\right)$, where
$\zeta = \left( \begin{smallmatrix} Q \\ \omega \end{smallmatrix} \right)$
and $A = \left( \begin{smallmatrix} a&b\\ b&c \end{smallmatrix} \right)$.
For cuts along K, the the constants describing the Gaussian were $a = 1109.0$ rlu$^{-2}$, $b = 0.0$ (rlu$\cdot$meV)$^{-1}$ and 
$c = 0.48$ meV$^{-2}$. For cuts along L, the the constants describing the Gaussian were 
$a = 579.7$ rlu$^{-2}$, $b = -20.0$ (rlu$\cdot$meV)$^{-1}$ and 
$c = 1.3$ meV$^{-2}$. This result was then convoluted with the model.

During the data reduction and analysis, the measured spin wave dispersion is binned and 
integrated over two directions and the remaining two directions plotted with the intensity 
given by the pixel color. To simulate this step, we integrated the calculated intensity over 
a volume of length $\pm$0.2 r.l.u. in the integrated 
directions and 0.05 r.l.u. (representing the bin size) in the remaining direction. 

\bibliography{YFeO3}

\end{document}